\documentclass[utf8]{frontiersFPHY} % for Physics and Applied Mathematics and Statistics articles

\setcitestyle{square} % for Physics and Applied Mathematics and Statistics articles
\usepackage{url,hyperref,lineno,microtype,subcaption}
\usepackage[onehalfspacing]{setspace}

\def\keyFont{\fontsize{8}{11}\helveticabold }
\def\firstAuthorLast{Wildy {et~al.}} %use et al only if is more than 1 author
\def\Authors{Conor Wildy\,$^{1,*}$, and Bozena Czerny\,$^{1}$}
\def\Address{$^{1}$Center for Theoretical Physics, Polish Academy of Sciences, Al. Lotnik\'{o}w 32/46, 02-668 Warsaw, Poland}
\def\corrAuthor{Conor Wildy}

\def\corrEmail{wildy@cft.edu.pl}

\begin{document}
\onecolumn
\firstpage{1}

\title[Mg~{\sc ii} emission and quasar inclination angle]{The relationship between Mg~{\sc ii} broad emission and quasar inclination angle} 

\author[\firstAuthorLast ]{\Authors} %This field will be automatically populated
\address{} %This field will be automatically populated
\correspondance{} %This field will be automatically populated

\extraAuth{}% If there are more than 1 corresponding author, comment this line and uncomment the next one.
%\extraAuth{corresponding Author2 \\ Laboratory X2, Institute X2, Department X2, Organization X2, Street X2, City X2 , State XX2 (only USA, Canada and Australia), Zip Code2, X2 Country X2, email2@uni2.edu}

\maketitle

\begin{abstract}

%%% Leave the Abstract empty if your article does not require one, please see the Summary Table for full details.
\section{}
Several observed spectral properties of quasars are believed to be influenced by quasar orientation. In this investigation we examine the effect of orientation on the Mg~{\sc ii} line located at 2798~\AA{} in a sample of 36 radio-loud quasars, with orientation angles having been obtained in a previous study using radio observations. We find no significant relationship between orientation angle and either Mg~{\sc ii} line full-width at half-maximum or equivalent width. The lack of correlation with inclination angle contradicts previous studies which also use radio data as a proxy for inclination angle and suggests the Mg~{\sc ii} emission region does not occupy a disk-like geometry. The lack of correlation with Mg~{\sc ii} equivalent width, however, is reported in at least one previous study. Although the significance is not very strong (86 percent), there is a possible negative relationship between inclination angle and Fe~{\sc ii} strength which, if true, could explain the Fe~{\sc ii} anti-correlation with [O~{\sc iii}] strength associated with Eigenvector 1. Interestingly, there are objects having almost edge-on inclinations while still exhibiting broad lines. This could be explained by a torus which is either clumpy (allowing sight lines to the central engine) or mis-aligned with the accretion disk.

\tiny
 \keyFont{ \section{Keywords:} Active galaxies, Accretion, Broad line region, Mg~{\sc ii}, Spectroscopy} %All article types: you may provide up to 8 keywords; at least 5 are mandatory.
\end{abstract}

\section{Introduction}

Active Galactic Nuclei (AGN) are luminous sources powered by accretion onto supermassive black holes at the centres of galaxies. Orientation angles are known to affect the observed properties of AGN. For example it is thought that Type 1 objects, which show broad lines in their unpolarized spectra, have a low inclination angle with respect to the accretion disk axis and therefore enable an unobscured view of the central engine, while Type 2 objects, which only exhibit narrow lines, are thought to be viewed at high inclination and therefore obscured by a dusty torus oriented such that it is co-planar with the accretion disk \citep{urry95}. This is supported by the detection of broad lines in polarized light from Type 2 objects \citep{miller90} and higher column density x-ray absorbers in Type 2 objects compared with Type 1 \citep{turner89}. It is plausible that the inclination angle can influence the broad line morphology within the group of Type 1 objects. Studies have already indicated this for H$\beta$ in multi-quasar samples, where the width of the broad emission line is dependent on the inclination angle such that at higher inclinations the line tends to be broader \citep{wills86,runnoe13}. Such a phenomenon is easily explained by assuming that the bulk of the broad line region (BLR) is located in a flattened disk-like structure, with the observed velocity of an individual BLR cloud ($v_{obs}$) being related to its intrinsic velocity ($v_{int}$) by ${\rm v_{obs}=v_{int}sin\theta{}_{i}}$, where $\theta{}_{i}$ is the inclination angle. 

Quasars are a luminous sub-category of AGN mostly exhibiting Type 1 features. Many thousands have their observed optical spectra catalogued within the Sloan Digital Sky Survey (SDSS) \citep{adelman08}, enabling statistical analysis of large samples of these objects. One such example relevant to quasar inclination angles was that of \citet{shen14} which examined the so-called Eigenvector 1 correlations originally described by \citet{boroson92}. This correlation is mainly due to the inverse relationship between the strengths of Fe~{\sc ii} and the optical narrow lines of [O~{\sc iii}], and also shows an inverse relationship between the width of H$\beta$ and its strength. The paper of \citet{shen14} showed that, in a sample of approximately 20\,000 objects, [O~{\sc iii}] strength is roughly constant at a given Fe~{\sc ii} strength, while the full-width at half-maximum (FWHM) of H$\beta{}$ varies significantly between objects. That paper concluded that this variation in FWHM was due to orientation, while the strength anti-correlation of Fe~{\sc ii} with [O~{\sc iii}] was not. It was described in \citet{risaliti11} that, since the emission of [O~{\sc iii}] is probably isotropic and scales with the quasar bolometric luminosity, then it is plausible that [O~{\sc iii}] EW tracks the quasar inclination angle, with higher EW indicating higher inclination angle. This was further supported in that paper by examination of the shape of the distribution of EW values. A recent study by \citet{bisogni17}, revealed that the Mg~{\sc ii} equivalent width (EW) was not dependent on [O{~\sc iii}] EW, suggesting that the emission of Mg~{\sc ii}, unlike the case of [O{~\sc iii}], was anisotropic. This is consistent with the postulate that BLR emission is located in a disk-like structure.

In this investigation we test the dependence of Mg~{\sc ii} width and strength on inclination angle. To do so we selected objects from those listed in \citet{kuzmicz12}, which have direct measurements of quasar orientation obtained from radio data and whose spectra are archived in the SDSS. We improve on the multi-component fit of the underlying continuum in that paper by including the Balmer continuum emission as well as the power-law continuum and broadened Fe~{\sc ii} templates, allowing us to obtain more accurate Mg~{\sc ii} EW measurements and profile shapes.

\section{Objects used in the quasar sample}

There are 91 quasars in the \citet{kuzmicz12} paper with optical and radio observations obtained from previous catalogues (see references therein), for which inclination angle measurements were obtainable. Inclination angles were calculated in that study using the follow equation:

\begin{equation}
\label{eqn:icang}
\theta{}_{i}=\rm{acos}\left(\frac{1}{\beta{}_{j}}\frac{\left(s-1\right)}{\left(s+1\right)}\right)\, ,
\end{equation}

\noindent where $s=(S_{j}/S_{cj})^{\frac{1}{\left(2-\alpha{}\right)}}$, $S_{j}$ and $S_{cj}$ are the peak flux density of the lobes appearing closer to and further from the core respectively and $\alpha{}$ is the spectral index which is assumed to be $\alpha{}$=$-$0.6 from \citet{wardle97}. The constant $\beta{}_{j}$ is the jet velocity which, in agreement with \citet{wardle97} and \citet{arshakian04}, was fixed at 0.6c.

Of the 91 quasars are included 43 objects which are categorized as \emph{Giant Radio Quasars} (GRQs), defined as objects having a radio structure $>$0.72~Mpc after applying the cosmological parameters of \citet{spergel03}. As that study was designed to reveal any differences between GRQs and the rest of the radio-loud quasar population, they also selected a comparison sample made up of 48 objects having a smaller radio structure. As their study found on average no statistically significant differences in quasar optical spectra between the two populations, we made no distinction between the two categories and treated all objects as one sample. Only objects having SDSS spectral coverage spanning the entire range 2300--3300 \AA{} in the quasar rest frame were subject to analysis, as this allows an adequate span to fit the underlying emission made up of the power-law continuum from the accretion disk, broadened Fe~{\sc ii} emission and Balmer continuum emission from the BLR. Finally, only 36 objects from the sample underwent scientific analysis due to: (i) five objects having no SDSS spectra; (ii) 30 objects not meeting the 2300--3300 \AA{} criterion; (iii) one object having heavily absorbed Mg~{\sc ii} emission; and (iv) 19 objects being unable to be fitted satisfactorily. These 36 objects and their spectral parameters measured in this study are listed in Table~\ref{tab:quasars}. All EW values are calculated with respect to the disk power-law continuum.

\begin{table*}
\begin{center}
\caption{List of quasars used for Mg~{\sc ii} analysis.}
\begin{tabular}{l l l l l l}
\hline SDSS name&Redshift&Inclination angle&Mg~{\sc ii} EW&Mg~{\sc ii} FWHM&Fe~{\sc ii} EW\\
 & &($^{\circ}$)&(\AA{})&(km s$^{-1}$)&(\AA{})\\
\hline
SDSS J005115.11$-$090208.5&1.259&55&53.0$\pm$6.6&8920$\pm$463&175$\pm$8\\
SDSS J021008.48$+$011839.6&0.870&63&41.1$\pm$4.7&4840$\pm$265&120$\pm$8\\
SDSS J075448.86$+$303355.1&0.796&87&54.1$\pm$5.4&5340$\pm$391&131$\pm$8\\
SDSS J080906.22$+$291235.4&1.481&28&33.5$\pm$3.5&5240$\pm$465&77.5$\pm$3.6\\
SDSS J081240.08$+$303109.4&1.313&71&43.8$\pm$19.2&2790$\pm$220&192$\pm$59\\
SDSS J081409.22$+$323731.9&0.843&72&87.8$\pm$5.3&3390$\pm$234&196$\pm$13\\
SDSS J090207.20$+$570737.8&1.592&79&32.2$\pm$4.2&5610$\pm$1130&117$\pm$8\\
SDSS J090429.62$+$281932.7&1.122&45&23.8$\pm$4.0&3420$\pm$244&77.4$\pm$3.6\\
SDSS J090649.98$+$083255.8&1.616&86&36.4$\pm$11.4&5300$\pm$757&115$\pm$10\\
SDSS J091858.15$+$232555.4&0.690&81&65.4$\pm$6.2&6670$\pm$688&81.7$\pm$5.9\\
SDSS J092425.02$+$354712.6&1.344&84&48.1$\pm$8.7&4980$\pm$1340&127$\pm$7\\
SDSS J094418.85$+$233119.9&0.989&83&63.2$\pm$7.4&5600$\pm$381&107$\pm$6\\
SDSS J095206.38$+$235245.2&0.971&89&34.1$\pm$9.9&3900$\pm$203&62.2$\pm$4.0\\
SDSS J095934.49$+$121631.5&1.091&73&40.1$\pm$5.5&5550$\pm$389&186$\pm$13\\
SDSS J100507.07$+$501929.8&2.016&76&26.3$\pm$21.6&4390$\pm$201&159$\pm$37\\
SDSS J100607.70$+$323626.1&1.026&80&85.2$\pm$7.5&3340$\pm$259&142$\pm$13\\
SDSS J102026.87$+$044752.0&1.134&61&48.3$\pm$10.9&6560$\pm$314&153$\pm$19\\
SDSS J102041.14$+$395811.2&0.830&58&77.3$\pm$29.5&9440$\pm$1390&247$\pm$16\\
SDSS J105636.25$+$410041.2&1.781&87&42.0$\pm$12.5&5510$\pm$534&189$\pm$29\\
SDSS J111023.84$+$032136.1&0.966&85&46.6$\pm$11.2&3090$\pm$329&90.8$\pm$14.6\\
SDSS J111858.62$+$382852.2&0.747&55&36.4$\pm$2.9&4920$\pm$467&53.3$\pm$8.4\\
SDSS J111903.28$+$385852.5&0.735&64&42.4$\pm$28.5&8790$\pm$191&49.7$\pm$8.3\\
SDSS J115139.68$+$335541.4&0.851&32&43.6$\pm$5.7&7200$\pm$612&79.9$\pm$5.1\\
SDSS J121701.37$+$101952.9&1.884&87&47.2$\pm$2.4&4110$\pm$330&103$\pm$7\\
SDSS J122925.53$+$355532.1&0.828&57&33.3$\pm$6.7&4080$\pm$191&149$\pm$14\\
SDSS J123604.51$+$103449.2&0.667&63&34.7$\pm$6.0&4890$\pm$791&39.2$\pm$8.0\\
SDSS J125607.67$+$100853.6&0.824&69&89.2$\pm$25.6&3030$\pm$151&140$\pm$20\\
SDSS J132106.65$+$374153.4&1.135&79&28.1$\pm$19.9&11600$\pm$837&85.4$\pm$9.4\\
SDSS J133411.70$+$550124.9&1.247&89&49.7$\pm$15.7&13100$\pm$292&54.7$\pm$7.0\\
SDSS J134034.70$+$423232.1&1.345&89&87.0$\pm$16.5&6630$\pm$687&243$\pm$20\\
SDSS J135817.60$+$575204.5&1.372&85&27.2$\pm$1.9&4760$\pm$276&109$\pm$4\\
SDSS J155002.00$+$365216.7&2.071&64&47.3$\pm$34.8&5920$\pm$346&192$\pm$18\\
SDSS J155729.93$+$330446.9&0.944&89&49.3$\pm$6.2&5640$\pm$204&147$\pm$9\\
SDSS J162229.93$+$353125.3&1.471&82&43.3$\pm$11.4&4550$\pm$621&133$\pm$10\\
SDSS J162336.45$+$341946.3&1.994&13&43.8$\pm$7.7&7360$\pm$485&0.103$\pm$11.548\\
SDSS J223458.73$-$022419.0&0.550&83&41.0$\pm$7.6&5300$\pm$629&41.0$\pm$4.7\\
\hline    
\end{tabular}
\label{tab:quasars}
\end{center}
\end{table*}

\section{Emission components in the spectral region of Mg~{\sc ii}}

\begin{figure}[h!]
\begin{center}
\includegraphics[width=10cm, angle=270]{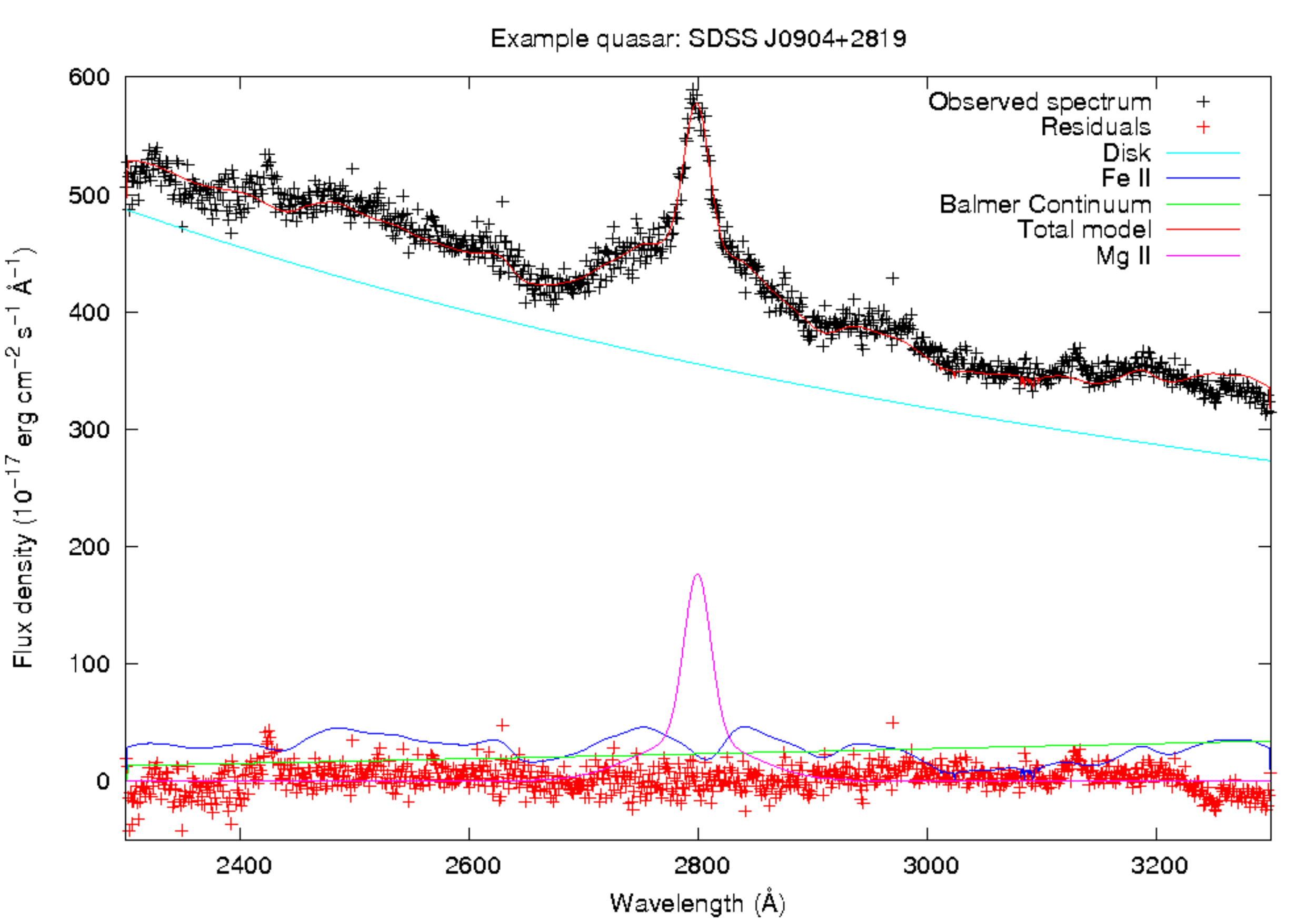}% This is a *.jpg file
\end{center}
\caption{Final fit to the spectrum of quasar SDSS J090429.62$+$281932.7 (black crosses) showing all four spectral components and the total model. Residuals are the difference between the total model and the observed datapoints.}\label{fig:qfit}
\end{figure}

The Mg~{\sc ii} line exists in a complex spectral region of the UV known as the "small blue bump" (SBB), whose underlying emission must be accurately reconstructed in order to correctly measure the Mg~{\sc ii} strength and width. The SBB consists of a BLR Balmer continuum and Fe~{\sc ii} emission made up of many overlapping broadened lines which, as they are blended, give the appearance of continuous emission. The SBB adds to the underlying power-law continuum to give the total continuous emission spanning the wavelengths occupied by the Mg~{\sc ii} line. This line itself is modelled using 1 or 2 Gaussian profiles, in the two profile case no significance is assigned to the properties of the individual components, they are used simply to generate the total profile from their sum. The modelled Fe~{\sc ii} emission is based on the template of \citet{tsuzuki06}. The calculation of the Balmer continuum was not performed in the fitting procedure of \citet{kuzmicz12}, hence our fitting method represents an important improvement over their technique. The Balmer continuum was calculated using the following equation obtained from \citet{grandi82}:  

\begin{equation}
\label{eqn:bccalc}
F_{\nu}^{\rm BC}=F_{\nu}^{\rm BE}e^{\left(-h-\nu{}_{\rm BE}\right)/\left(kT_{e}\right)}\, ,
\end{equation}

\noindent where $F_{\nu}^{\rm BC}$ is the BC flux at frequency $\nu{}$, $F_{\nu}^{\rm BE}$ is the flux of the BC at the Balmer edge (3646~\AA{}), $T_{e}$ is the electron temperature, $k$ is the Boltzmann constant and $h$ is the Planck constant. The fitting of all three SBB components was performed in a four-step manner for each quasar. First, the power-law continuum was fitted to relatively line-free regions outside the approximate extent of the SBB (less than 2300~\AA{} and greater than 3700~\AA{}) using the \emph{specfit} software within the Image Reduction and Analysis Facility ({\sc iraf}). Second, the Fe~{\sc ii} template and Balmer continuum were fitted to spectral regions spanning 2300--2600~\AA{} and 3000--3300~\AA{} using chi-square minimization, after subtraction of the power-law continuum. These wavelength ranges were used as they are neighbouring bands on either side of the Mg~{\sc ii} line. The electron temperature of the plasma generating the Balmer continuum was treated as a free parameter and allowed to vary between 4000 and 20\,000~K in steps of 1000~K, as was the Balmer edge flux which varied in steps of 0.01 percent of the continuum subtracted quasar flux. For the Fe~{\sc ii} template, the free parameters were Gaussian broadening of between 0 and 3000~km~s$^{-1}$ in steps of 1000~km~s$^{-1}$ and the normalization, which varied in steps of 0.01 percent of the continuum subtracted quasar flux. Third, the Mg~{\sc ii} model was fitted to each quasar spectrum with the power-law and SBB continuum subtracted, using \emph{specfit}. Fourth, all four components were simultaneously re-fitted to the region 2300--3300~\AA{} using \emph{specfit}, with only normalizations allowed to vary. An image of an example resultant fit is shown in Figure~\ref{fig:qfit}.

\section{Analysis}

\subsection{Relationship between Mg~{\sc ii} width and quasar inclination angle}

\begin{figure}[h!]
\begin{center}
\includegraphics[width=10cm, angle=270]{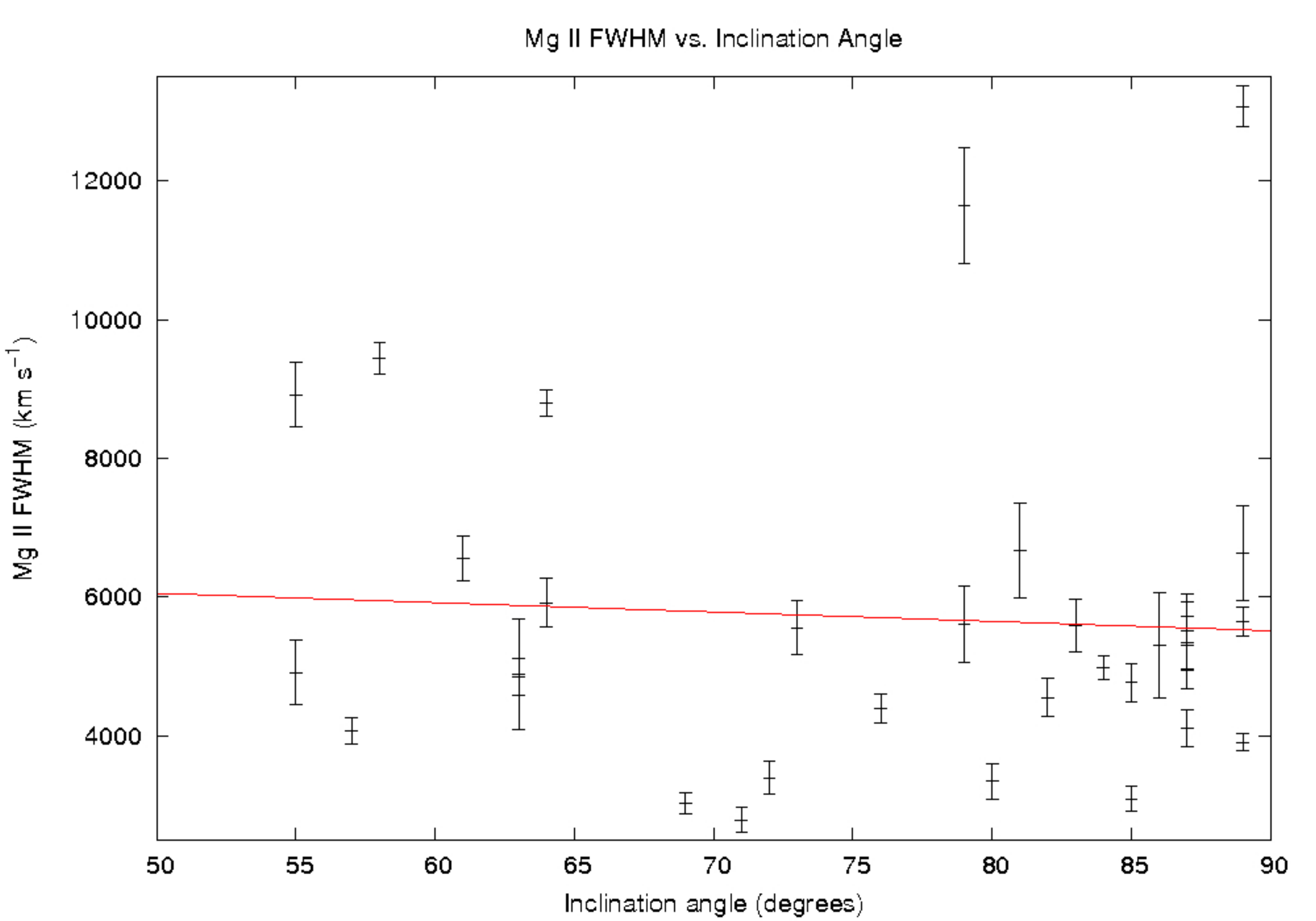}% This is a *.jpg file
\end{center}
\caption{Quasar Mg~{\sc ii} FWHM plotted against inclination angle, with vertical bars at each point indicating FWHM 1$\sigma$ error range for each object. A best-fitting trend line (red) shows a small negative gradient, however a Pearson test indicates no correlation.}\label{fig:mgfwhm}
\end{figure}

A Pearson test for correlation between the Mg~{\sc ii} FWHM and the quasar inclination angle for the 36 objects was performed. This found a probability of 38 percent that there is an actual correlation between the two variables, indicating that the data is consistent with the null hypothesis (no correlation). Each quasar's FWHM and inclination angle, along with the best-fitting linear trend-line, is illustrated in Figure~\ref{fig:mgfwhm}.

\subsection{Relationship between equivalent width and inclination angle}

\begin{figure}[h!]
\begin{center}
\includegraphics[width=20cm, angle=270]{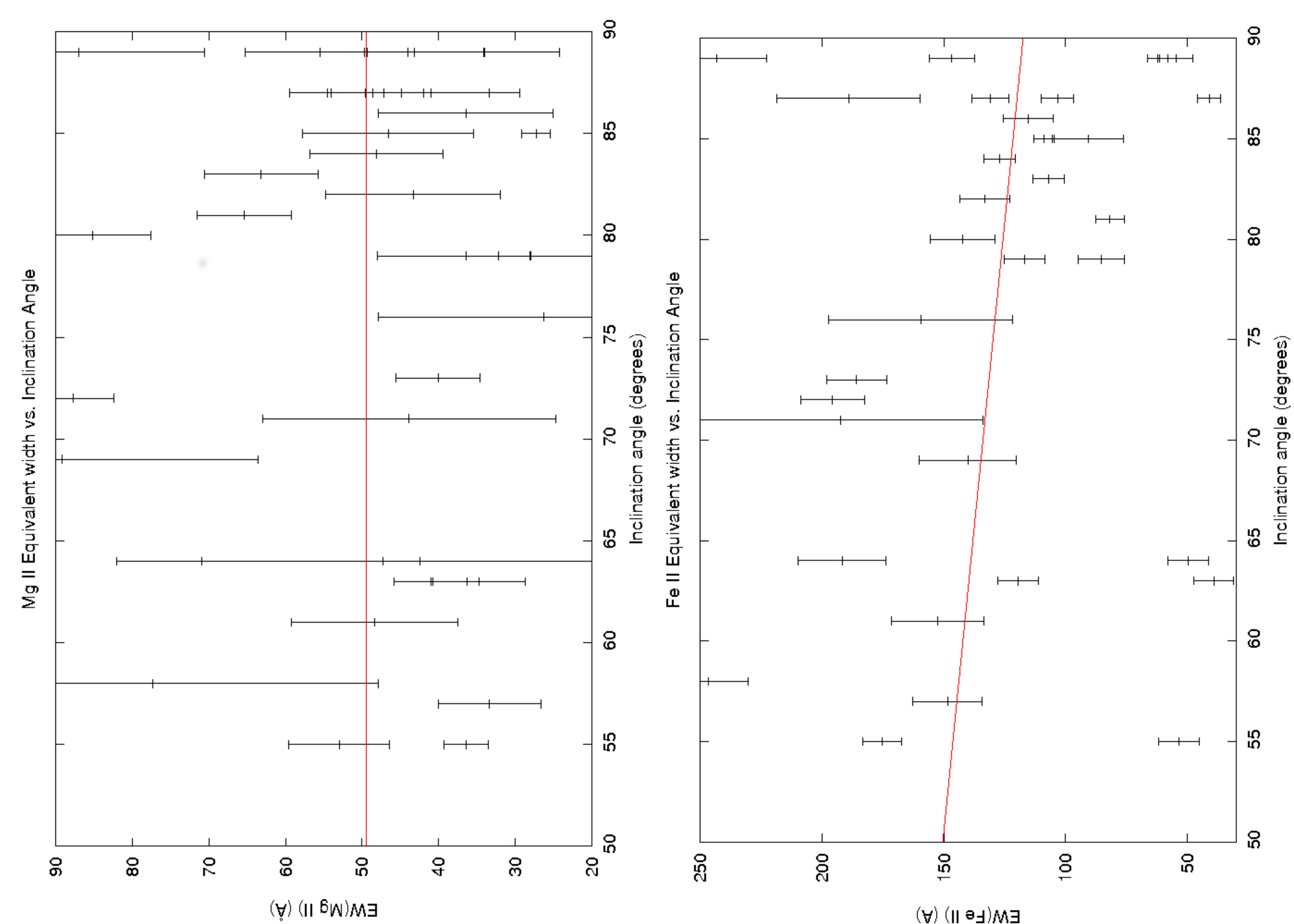}% This is a *.jpg file
\end{center}
\caption{Quasar Mg~{\sc ii} EW  (top panel) and Fe~{\sc ii} EW (bottom panel) plotted against inclination angle along with the respective best-fitting trend lines (red). Vertical bars at each point indicate EW 1$\sigma$ error range for each object. For Fe~{\sc ii} the correlation likelihood is higher (86 percent) when compared to that of Mg~{\sc ii} (38 percent).}\label{fig:mgfeew}
\end{figure}

Similar to the procedure for Mg~{\sc ii} FWHM, the EW of the line was tested across all 36 quasars using a Pearson test for correlation. This provided a value of 38 percent likelihood of correlation. Again, this is consistent with the null hypothesis of no correlation. The Fe~{\sc ii} EW was also measured over the same wavelength span as the Mg~{\sc ii} line, revealing that there may be a hint of a correlation between the Fe~{\sc ii} EW and inclination angle since the Pearson test indicates an 86 percent likelihood of correlation, much higher than for the other cases mentioned so far. If such a correlation does in fact exist, it is likely to be a negative relationship given that the correlation coefficient is negative. The relationship of both Mg~{\sc ii} and Fe~{\sc ii} EW as a function of inclination angle is illustrated in Figure~\ref{fig:mgfeew}.

\subsection{Broad line detection at large inclination angles}

\begin{figure}[h!]
\begin{center}
\includegraphics[width=10cm, angle=270]{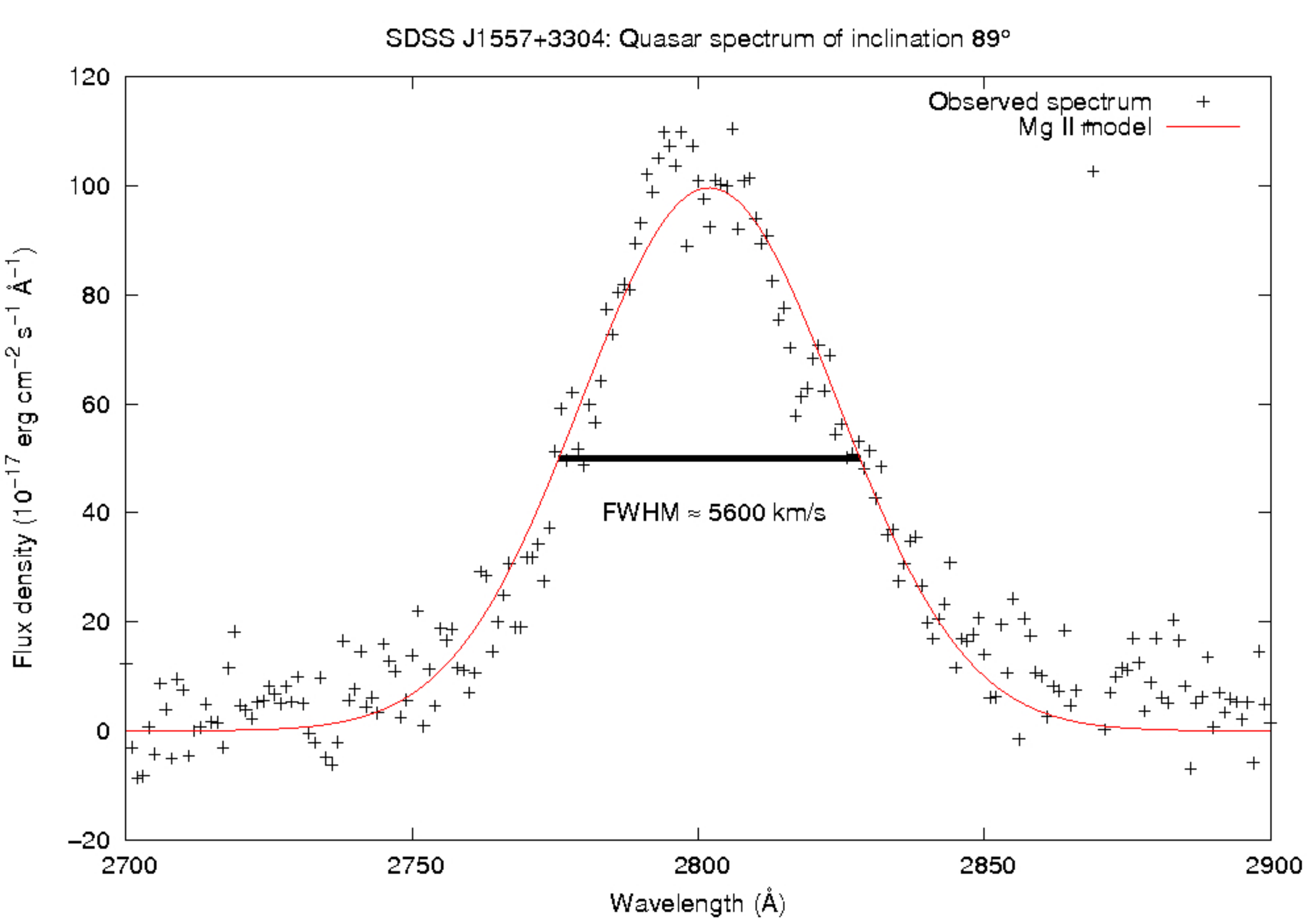}% This is a *.jpg file
\end{center}
\caption{Mg~{\sc ii} broad line emission as it appears in the quasar SDSS J155729.93$+$330446.9. The black crosses indicate the quasar flux after the disk power-law and the SBB emission were subtracted. The red line indicates the model Mg~{\sc ii} emission which, at 5600~km~s$^{-1}$, easily meets the definition of a broad line.}\label{fig:quasar89}
\end{figure}

One of the most interesting findings from the sample is the detection of broad emission lines at almost edge-on inclinations. This is somewhat surprising as it could be expected that at high inclinations the dusty torus obscures the view to the central source of the quasar. There are several examples of such objects in our sample, one of which is shown in Figure~\ref{fig:quasar89}.

\section{Discussion}

The fact that we do not detect a correlation between the Mg~{\sc ii} broad line width in our sample and the inclination angle is a somewhat surprising result which contradicts reports in other studies. For example, \citet{aars05} found a correlation between the radio-derived inclination angle and the Mg~{\sc ii} FWHM, which is in accordance with what would be expected if the broad line region, including the Mg~{\sc ii} emitting gas, were located in a flattened-disk geometry. A flattened BLR is further supported by the study of \citet{kimball11} which found no correlation between the Mg~{\sc ii} EW and the ratio of radio core flux to lobe flux, the latter being an inclination indicator. If instead a dependence between these two values was to have been found, this would imply that the broad line emission is substantially more isotropic than that of the accretion disk and therefore that the BLR cannot have a disk geometry. In the case of [O~{\sc iii}] the line emission is isotropic and therefore its EW could potentially track the inclination angle \citep{risaliti11}.

Like \citet{kimball11} our findings also indicate a lack of correlation between Mg~{\sc ii} EW and inclination angle. Considering the lack of correlation of either line width or EW with inclination in our study, it is possible that the BLR exists in an intermediate geometry between an isotropically emitting sphere and an anisotropically emitting thin-disk. Possibilities for such geometries include a thick disk, a distorted disk, or some combination of both. This could result in only a weak relationship between inclination and line width, rendering it undetectable in our sample size, while still presenting no strong evidence of isotropic emission from the relationship between inclination angle and line EW. This could explain the correlation between [O~{\sc iii}] EW and broad line width seen in \citet{risaliti11}, as their sample size is much larger than ours, consisting of approximately 6000 SDSS quasars.

There is no strong indication of a correlation between Fe~{\sc ii} EW and inclination (such a conclusion would require a confidence $>$95 percent), however it is stronger than the others, at 86 percent. This correlation, if it exists, is negative. If the [O~{\sc iii}] emission is an accurate indicator of inclination, then this would support the findings of \citet{bisogni17} and may be a contributor to the anti-correlation between [O~{\sc iii}] and Fe~{\sc ii} strength found from the Eigenvector 1 relationship \citep{boroson92}. A possible physical explanation for this, if the anti-correlation is true, is the Fe~{\sc ii} emission being located in a disk which is even thinner than the accretion disk \citep{bisogni17}, however this is speculative. It is not possible to accurately measure the Fe~{\sc ii} line widths using our method as there are only three values of broadening used and in many cases more than one value gives an acceptable fit to the spectrum.

The fact that broad lines are visible at high inclinations requires an explanation, since, in the common unification explanation for different AGN types, Type 1 Seyferts/quasars should have an inclination angle lower than approximately 45$^{\circ{}}$. Broad lines are only visible in Type 2 objects in polarized light \citep{antonucci85}. In fact, most of the objects studied here have inclination angles greater than 45$^{\circ{}}$. If the torus is not smooth but is instead a clumpy structure, then this may allow a line-of-sight toward the central engine in the objects observed. In such a case the probability of detecting a broad line doesn't reach zero even if the inclination angle tends towards 90$^{\circ{}}$. Gas inflow models have given support to a clumpy nature of the dusty torus \citep{netzer15}, providing evidence that this structure is possible. A further possibility is that of the torus being misaligned with the accretion disk in these cases. It should be noted in any case that the \citet{kuzmicz12} selection process necessarily found quasars at high inclination angle, since the selection criteria required objects to have both a large angular size and to be lobe-dominated. Quasars of this kind are very rare in the general quasar population \citep{devries06} and so our sample objects may have unusual properties compared to those in previous quasar samples compiled to determine the effects of inclination angle on spectral morphology. 

\section*{Conflict of Interest Statement}

The authors declare that the research was conducted in the absence of any commercial or financial relationships that could be construed as a potential conflict of interest.

\section*{Author Contributions}

CW was responsible for obtaining and analyzing the data from the SDSS, as well as interpreting the results. BC devised and supervised the project. 

\section*{Funding}

This project was supported by the Polish Funding Agency National Science Centre, project 2015/17/B/ST9/03436/ (OPUS 9).

\section*{Acknowledgments}
This work is based on spectroscopic observations made by the Sloan Digital Sky Survey. Funding for the Sloan Digital Sky Survey IV has been provided by the Alfred P. Sloan Foundation, the U.S. Department of Energy Office of Science, and the Participating Institutions.

\bibliographystyle{frontiersinHLTH&FPHY} % for Health, Physics and Mathematics articles
\bibliography{bib_cw}

%%% Make sure to upload the bib file along with the tex file and PDF
%%% Please see the test.bib file for some examples of references

%%% Please be aware that for original research articles we only permit a combined number of 15 figures and tables, one figure with multiple subfigures will count as only one figure.
%%% Use this if adding the figures directly in the mansucript, if so, please remember to also upload the files when submitting your article
%%% There is no need for adding the file termination, as long as you indicate where the file is saved. In the examples below the files (logo1.jpg and logos.jpg) are in the Frontiers LaTeX folder
%%% If using *.tif files convert them to .jpg or .png
%%%  NB logo1.jpg is required in the path in order to correctly compile front page header %%%

%%% If you are submitting a figure with subfigures please combine these into one image file with part labels integrated.
%%% If you don't add the figures in the LaTeX files, please upload them when submitting the article.
%%% Frontiers will add the figures at the end of the provisional pdf automatically
%%% The use of LaTeX coding to draw Diagrams/Figures/Structures should be avoided. They should be external callouts including graphics.

\end{document}